\documentclass[useAMS,usenatbib]{mn2e}
\usepackage{float,multicol,epsfig}
\usepackage{amssymb}

\title[The effect of forbidden transitions on H and 
He recombination]
{The effect of forbidden transitions on cosmological hydrogen 
and helium recombination}

\author[W. Y. Wong and D. Scott]{Wan Yan Wong\thanks{E-mail:
wanyan@phas.ubc.ca}
and Douglas Scott\thanks{E-mail: dscott@phas.ubc.ca} \\
Department of Physics and Astronomy, University of British Columbia,
 6224 Agricultural Rd., Vancouver, BC, V6T 1Z1, Canada}

\begin{document}
\date{2006 December}

\pagerange{\pageref{firstpage}--\pageref{lastpage}} \pubyear{2006}

\maketitle

\label{firstpage}
\begin{abstract}
More than half of the atoms in the Universe recombined via
forbidden transitions, so that accurate treatment of the forbidden
channels is important in order to follow the cosmological recombination
process with the level of precision required by future microwave 
anisotropy experiments.
We perform a multi-level calculation of the recombination of hydrogen
(H) and helium (He) with the addition of the \mbox{$2^3$P$_1$--$1^1$S$_0$} 
spin-forbidden transition for neutral helium (He\,{\small I}),
plus the $n$S--1S and $n$D--1S two-photon transitions for H (up to $n=40$)
and among singlet states of He\,{\small I} ($n \leq 10$ and $\ell \leq 7$).  
The potential importance of such transitions was first proposed by 
\citet{dub05} using an effective three-level atom model.  Here, we relax 
the thermal equilibrium assumption among the higher excited states to 
investigate the effect of these extra forbidden transitions on the 
ionization fraction $x_{\rm e}$ and the Cosmic Microwave Background (CMB) 
angular power spectrum $C_\ell$.  The spin-forbidden transition 
brings more than a percent change in $x_{\rm e}$.  The two-photon
transitions may also give non-negligible effects, but currently accurate
rates exist only for $n \leq 3$. We find that changes in both $x_{\rm e}$
 and $C_\ell$ would be at about the percent level with the approximate rates 
given by \citet{dub05}.  However, the two-photon rates from 3S to 1S 
and 3D to 1S of H appear to have been overestimated, and our best 
numerical calculation puts the effect on $x_{\rm e}$ and $C_\ell$ at below the 
percent level.  Nevertheless, we do not claim that we have the definite answer, 
since several issues remain open; sub-percent level computation of the $C_\ell$s
requires improved calculations of atomic transition rates as well
as increasingly complex multi-level atom calculations.
\end{abstract}

\section{Introduction}
The release of the third year data from the Wilkinson Microwave Anisotropy 
Probe ({\sl WMAP}) has further improved the precision with which we can
constrain the cosmological parameters from the shape of 
the Cosmic Microwave Background (CMB) anisotropies $C_\ell$~\citep{spergel06}.
The {\sl Planck\/} satellite, scheduled for launch in 2008~\citep{planck06}, 
will provide even higher precision $C_\ell$ values and data down 
to smaller angular scales ($\ell \lesssim 2500$).  Higher precision in 
the observations requires increased accurarcy
from the theoretical calculations, in order for the correct cosmological parameters 
to be extracted.  It now seems crucial to obtain the $C_\ell$s down to 
at least the 1 percent level over a wide range of $\ell$.

{\sc cmbfast}~\citep{cmbfast} is the most commonly used Boltzmann code for calculating 
the $C_\ell$s, and it gives consistent results with other independent codes
\citep[see][and references therein]{seljak03}.  The dominant 
uncertainty in obtaining accurate $C_\ell$s comes from details in the physics
of recombination, for example, the `fudge factor' in the {\sc recfast} 
routine~\citep{seager99, seager00}.
Calculations of cosmological recombination were first published by~\citet{peebles68}
and~\citet{zks68}. \citet{seager00} presented the most detailed multi-level 
calculation and introduced a fudge factor to reproduce the results within an effective
three-level model.  Although the multi-level calculation 
already gives reasonable accuracy, the required level of accuracy continues to
increase, so that today any effect which is $\sim$\,1 per cent over a range of 
multipoles is potentially significant.  Several modifications have been recently 
suggested to give per cent level changes in the ionization fraction 
and/or the $C_\ell$s (see Section 4 for details).
Most of these modifications have been calculated only with an effective 
three-level code, and so the results may be different in the 
multi-level calculation, since there is no thermal equilibrium assumed 
between the upper states.  Here we want to focus on one of these modifications, 
namely the extra forbidden transitions proposed by \citet{dub05}, which
we study using a multi-level code.  

In the standard calculations of recombination, one
considers all the resonant transitions, but only one forbidden transition, 
which is the 2S--1S two-photon transition, and this can be included 
for both H and He.  \citet{dub05} suggested that one should also include 
the two-photon transitions from higher excited S and D states to the 
ground state for H and He\,{\small I}, and 
also the spin-forbidden transition between the triplet $2^3$P$_1$ and 
singlet ground state $1^1$S$_0$ for He\,{\small I}.  They showed 
that the recombination of both H and He\,{\small I} sped up in the 
three-level atom model.  The suggested level of change is large enough 
to bias the determination of the cosmological parameters~\citep{lewis06}. 

The purpose of this paper is to investigate the effect of the extra forbidden 
transitions suggested by~\citet{dub05} in the multi-level atom model without
assuming thermal equilibrium among the higher excited states.  The outline 
of the paper is as follows.  In Section 2 we will describe details of the 
rate equations in our numerical model.  In Section 3 we will present results 
on the ionization fraction $x_{\rm e}$ and the anisotropies $C_\ell$, and 
assess the importance of the addition of the forbidden transitions. Other 
possible improvements of the recombination calculation will be discussed in Section
4.  And finally in Section 5 we will present our conclusions.
%
\section{Model}
In this paper we follow the formalism of the multi-level calculation
performed by \citet{seager00}.  We consider 100 levels for H\,{\small I},
 103 levels for He\,{\rm I}, 10 levels for He\,{\small II}, 1 level for
He\,{\small III}, 1 level for the electrons and 1 level for the protons.
For H, we only consider discrete $n$ levels and assume that the angular sub-levels
($\ell$-states) are in statistical equilibrium within a given shell.  
For both He\,{\small I} and He\,{\small II}, we consider all the 
$\ell$-states separately. The multilevel He\,{\rm I} atom includes all
states with $n \leq 10$ and $\ell \leq 7$.  Here we give a summary of 
the rate equations for the number density of each energy level $i$, 
and the equation for the change of matter temperature $T_{\rm M}$.  
The rate equation for each state with respect to redshift $z$ is
{\setlength\arraycolsep{2pt}
\begin{eqnarray}
 (1+z) \frac{dn_i}{dz} &=& -\frac{1}{H(z)} \times \nonumber \\
   && \left[ \left( n_{\rm e} n_{\rm c} R_{{\rm c}i} - n_i R_{i {\rm c}} \right) 
	+ \sum^N_{j=1} \Delta R_{ji} \right] + 3 n_i , 
\end{eqnarray}} 
\\
where $n_i$ is the number density of the $i$th excited atomic state, $n_{\rm e}$
is the number density of electrons, and $n_{\rm c}$ is the number density
of continuum particles such as a proton, He\,{\small II}, or He\,{\small III} ion. 
Additionally $R_{{\rm c}i}$ is the photo-recombination rate, $R_{i {\rm c}}$ is the 
photo-ionization rate, $\Delta R_{ji}$ is the net bound-bound rate for 
each line transition, and $H(z)$ is the Hubble parameter.
We do not include the collisional rates, as they have been shown to be
negligible~\citep{seager00}. 

For He\,{\rm I}, we update the atomic data for the energy 
levels~\citep{MWD06}, the oscillator strength for resonant 
transitions~\citep[][in preparation]{drake06} and the photo-ionization 
cross-section spectrum.   We use the photo-ionization cross-section given by 
\citet{hs98} for $n$\,$\leq$\,10 and $\ell$\,$\leq$\,4, and adopt 
the hydrogenic approximation for states with $\ell$\,$\geq$\,5~\citep{sh91}.
It is hard to find published accurate and complete data for the photo-ionization
cross-section of He\,{\rm I} with large $n$ and $\ell$.  For example, 
a recent paper by \citet{bauman05} claimed that they had calculated 
the photo-ionization cross-section up to $n$\,$=$\,27 and $\ell$\,$=$\,26, 
although, no numerical values were provided.

The rate of change of matter temperature with respect to redshift is 
\begin{equation}
(1+z) \frac{d T_{\rm M}}{dz} = \frac{8 \sigma_{\rm T} U}{3 H(z) m_{\rm e} c}
\frac{n_{\rm e}}{n_{\rm e} + n_{\rm H} + n_{\rm He}} (T_{\rm M} - T_{\rm R})
+ 2T_{\rm M},
\end{equation}
where $T_{\rm R}$ is the radiation temperature, $n_{\rm He}$
is the total number density for helium, $m_{\rm e}$ is the
electron mass, $c$ is the speed of light,  $U= a_{\mathrm{R}}
T_{\mathrm{R}}^4$, $a_{\rm R}$ is the radiation constant and
$\sigma_{\mathrm{T}}$ is the Thompson scattering cross-section.

\citet{seager00} considered all the resonant transitions and only one 
forbidden transition, namely the 2S--1S two-photon transition, 
in the calculation of each atom, (for He\,{\small I}, 
2S\,$\equiv 2^{1}$S$_{0}$ and 1S\,$\equiv 1^{1}$S$_{0}$).  
The 2S--1S two-photon transition rate is given by 
\begin{equation}
\Delta R_{\rm 2S \rightarrow 1S} = \Lambda_{2 \rm S} \left( 
n_{2\rm S} - n_{1 \rm S} \frac{g_{2 \rm S}}{g_{1 \rm S}}
e^{-h_{\rm P} \nu_{2\rm S-1\rm S}/k_{\rm B} T_{\rm M}}\right), 
\end{equation}
where $\Lambda_{2 \rm S}$ is the spontaneous rate of the corresponding
 two-photon transition, $\nu_{2\rm S-1\rm S}$ is the frequency 
between levels 2S and 1S, $g_i$ is the degeneracy of the energy level
$i$, $h_{\rm P}$ is Planck's constant and $k_{\rm B}$ is Boltzmann's constant. 

Here we include the following extra forbidden transitions, which
were first suggested by \citet{dub05}.  The first ones are 
the two-photon transitions from $n$S and $n$D to 1S for H, plus 
$n^{1}$S$_{0}$ and $n^{1}$D$_{2}$ to $1^{1}$S$_{0}$ for He\,{\small I}.
For example, for H, we can group together the $n$S and $n$D states coming from
 the same level, so that we can write the two-photon transition rate as  
{\setlength\arraycolsep{1pt}
\begin{eqnarray}
\Delta R^{\rm H}_{n {\rm S} + n {\rm D}  \rightarrow 1{\rm S}} &=& 
\Lambda^{\rm H}_{n {\rm S} + n {\rm D}} \times \nonumber \\
&& 
\left( n_{n {\rm S} + n {\rm D}} - n_{1 \rm S} 
\frac{g_{n {\rm S} + n {\rm D}}}{g_{1 \rm S}}
e^{-h_{\rm P} \nu_{n1}/k_{\rm B} T_{\rm M}}\right).
\end{eqnarray}}
\\
Here $n$ (without a subscript) is the principle quantum number of the state,
$n_{n {\rm S} + n {\rm D}}$ is the total number density of the excited
atoms in either the $n$S or $n$D states,
and $\Lambda^{\rm H}_{n {\rm S} + n {\rm D}}$ is the effective  
spontaneous rate of the two-photon transition from $n {\rm S} + n {\rm D}$ to 1S, 
which is approximated by the following formula~\citep{dub05}:
\begin{equation}
\Lambda^{\rm H}_{n {\rm S} + n {\rm D}} =
\frac{54 \Lambda^{\rm H}_{2 \rm S} }{g_{n{\rm S} + n {\rm D}}}
\left( \frac{n-1}{n+1} \right)^{2n} \frac{11 n^2 - 41}{n} \, ,
\end{equation}
where $ \Lambda^{\rm H}_{2 \rm S}$ is equal to 8.2290\,s$^{-1}$~\citep{goldman89,santos98}. 
The latest value of $ \Lambda^{\rm H}_{2 \rm S}$ is equal to 
 8.2206\,s$^{-1}$~\citep{labzowsky05} and does not bring any 
noticeable change to the result.  Here $g_{n{\rm S} + n {\rm D}}$
is equal to 1 for $n$\,$=$\,2, and 6 for $n$\,$\ge$\,3. 
This spontaneous rate is estimated by considering only the non-resonant
two-photon transitions through one intermediate state $n$P.  
\citet{dub05} ignored the resonant two-photon transition contributions,
 since the escape probability of these emitted photons is very low.
The above formula for $\Lambda^{\rm H}_{n {\rm S} + n {\rm D}}$ is 
valid up to $n$\,$\simeq$\,$40$, due to the dipole approximation used,  
although it is not trivial to check how good this approximate rate is.  Besides the 
2S--1S two-photon rate, only the non-resonant two-photon rates from 3S to 1S and
3D to 1S are calculated accurately and available in the literature.
\citet{ctsc86} evaluated $\Lambda^{\rm H}_{\rm 3S}$ and 
$\Lambda^{\rm H}_{\rm 3D}$ by including the non-resonant
transitions through the higher-lying intermediate $n$P states ($n$\,$\ge$\,$4$),
 which are equal to 8.2197\,s$^{-1}$ 
and 0.13171\,s$^{-1}$, respectively.  These values were confirmed 
by \citet{fsm88} and agreed to three significant figures.  Using these
values, we find that $\Lambda^{\rm H}_{n {\rm S} + n {\rm D}}$ is equal 
to 1.484\,s$^{-1}$, which is an order of magnitude smaller than 
the value from the approximated rate coming from equation~(5).  The 
approximation given by \citet{dub05} therefore seems to be an 
overestimate.  This leads us instead 
to consider a scaled rate $\tilde{\Lambda}^{\rm H}_{n {\rm S} + n {\rm D}}$,
which is equal to $\Lambda^{\rm H}_{n {\rm S} + n {\rm D}}$ multiplied
by a factor to bring the approximated two-photon rates of H (equation~(5)) 
with $n$\,$=$\,3 into agreement with the numerical value given above, i.e.
\begin{equation}
\tilde{\Lambda}^{\rm H}_{n {\rm S} + n {\rm D}} =
0.0664 \ \Lambda^{\rm H}_{n {\rm S} + n {\rm D}}.
\end{equation}

Note that the use of the non-resonant rates is an approximation.
The resonant contributions are suppressed in practice because of
optical depth effects, and in a sense some of these contributions
are already included in our multi-level calculation.  Nevertheless,
the correct way to treat these effects would be in a full 
radiative transfer calculation, which we leave for a future study.
For He\,{\small I}, we treat $n^{1}$S$_{0}$ and  $n^{1}$D$_{2}$ 
separately and use equation~(3) for calculating the transition rates.  
The spontaneous rate $\Lambda^{\rm HeI}_{n {\rm S}/n {\rm D}} $ 
is estimated by \citet{dub05} by assuming a similar form to 
that used for $\Lambda^{\rm H}_{n {\rm S} + n {\rm D}}$:
\begin{equation}
\Lambda^{\rm HeI}_{n {\rm S}/n {\rm D}} =
\frac{1045 A^{\rm HeI} }{g_{n{\rm S} + n {\rm D}}}
\left( \frac{n-1}{n+1} \right)^{2n} \frac{11 n^2 - 41}{n} \, ,
\end{equation}
where $A^{\rm HeI}$ is a fitting parameter~(which is still uncertain 
 both theoretically and experimentally). According to 
\citet{dub05}, resonable values of $A$ range from 10 to 12\,s$^{-1}$, 
and we take $A$\,$=$\,11\,s$^{-1}$ here.  In our calculation, we include 
these extra two-photon rates up to $n$\,$=$\,40
for H and up to $n=10$ for He\,{\small I}.

The other additional channel included is the spin-forbidden 
transition between the triplet $2^3$P$_1$ and singlet $1^1$S$_0$ states in 
He\,{\rm I}. This is an intercombination/semi-forbidden electric-dipole 
transition which emits a single photon and therefore we can calculate 
the corresponding net rate by using the bound-bound resonant rate 
expression, i.e. 
{\setlength\arraycolsep{2pt}
\begin{eqnarray}
\Delta R_{2^3{\rm P}_1-1^1{\rm S}_0} &=& p_{2^3{\rm P}_1, 1^1{\rm S}_0} \times
 \nonumber \\
&& \left( n_{2^3{\rm P}_1} R_{2^3{\rm P}_1, 1^1{\rm S}_0}
- n_{2^1{\rm S}_0} R_{1^1{\rm S}_0, 2^3{\rm P}_1}
\right),
\end{eqnarray}}
\\ where 
{\setlength\arraycolsep{2pt}
\begin{eqnarray}
&&  R_{2^3{\rm P}_1,1^1{\rm S}_0} = 
A_{2^3{\rm P}_1, 1^1{\rm S}_0} + B_{2^3{\rm P}_1, 1^1{\rm S}_0} \bar{J}, \\
&&  R_{1^1{\rm S}_0,2^3{\rm P}_1} = B_{1^1{\rm S}_0,2^3{\rm P}_1} \bar{J}, \\
&& p_{2^3{\rm P}_1, 1^1{\rm S}_0} = \frac{1-e^{-\tau_{\rm s}}}{\tau_{\rm s}}, 
\quad \rm{with} \\
&& \tau_{\rm s} = \frac{A_{2^3{\rm P}_1, 1^1{\rm S}_0}\lambda^3_{2^3{\rm P}_1, 
1^1{\rm S}_0}
}{8 \pi H(z)} \left[ \frac{g_{2^3{\rm P}_1} }
{g_{ 1^1{\rm S}_0}} n_{1^1{\rm S}_0} - n_{2^3{\rm P}_1}  \right] .
\end{eqnarray}} 
\\
Here $A_{2^3{\rm P}_1, 1^1{\rm S}_0}$, $B_{2^3{\rm P}_1, 1^1{\rm S}_0}$ 
and $B_{1^1{\rm S}_0,2^3{\rm P}_1}$ are the
Einstein coefficients, $p_{2^3{\rm P}_1, 1^1{\rm S}_0}$ is the
Sobolev escape probability, $\tau_{\rm s}$ is the Sobolev optical 
depth~\citep[see][and references therein]{seager00}, 
$\lambda_{2^3{\rm P}_1, 1^1{\rm S}_0}$ is the wavelength of the energy 
difference between states $2^3{\rm P}_1$ and $1^1{\rm S}_0$,
and $\bar{J}$ is the blackbody intensity with temperature $T_{\rm R}$. 

This $2^3$P$_1$--$1^1$S$_0$ transition is not the lowest transition 
between the singlet and the triplet states.  
The lowest one is the forbidden magnetic-dipole transition 
between $2^3$S$_1$ and $1^1$S$_0$, 
with Einstein coefficient \mbox{$A_{2^3{\rm S}_1,1^1{\rm S}_0}$ =  
$1.73 \times 10^{-4}$\,s$^{-1}$}~\citep{lin77}.  However, this is much smaller 
than \mbox{$A_{2^3{\rm P}_1,1^1{\rm S}_0} 
= 177.58$\,s$^{-1}$}~\citep[][in preparation]{lach01,drake06},
so this transition can be neglected.  Note that \citet{dub05} used an older value of 
\mbox{$A_{2^3{\rm P}_1,1^1{\rm S}_0} = 233$\,s$^{-1}$} \citep{lin77} 
in their calculation. 

We use the Bader-Deuflhard semi-implicit numerical integration 
scheme~\citep[see Section 16.6 in][]{nr} to solve the above rate equations.
All the numerical results are carried out using the $\Lambda$CDM model with
cosmological parameters: $\Omega_{\rm B}$\,$=$\,0.04; $\Omega_{\rm CDM}$\,$=$\,0.2;
$\Omega_{\Lambda}$\,$=$\,0.76; $\Omega_{\rm K}$\,$=$\,0; 
$Y_{\rm p}$\,$=$\,0.24; $T_0$\,$=$\,2.725\,K
and $h$\,=\,0.73~\citep[consistent with those in][]{spergel06}. Here $Y_{\rm p}$ is
the primordial He abundance and $T_0$ the present background
temperature.
\section{Results}
\subsection{Change in ionization fraction}
\begin{figure}
\centering
\vspace*{7cm}
\leavevmode
\includegraphics{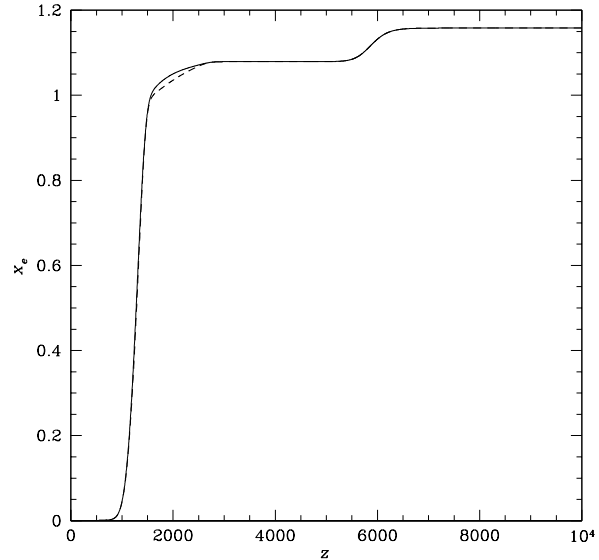}
\caption{The ionization fraction $x_{\rm e}$ as a function of redshift $z$.
The solid line is calculated using the original multi-level code of
\citet{seager00}, while the dashed line includes all the extra forbidden 
transitions discussed here.}
\label{graphxe}
\end{figure}
\begin{figure}
\centering
\vspace*{7cm}
\leavevmode
\includegraphics{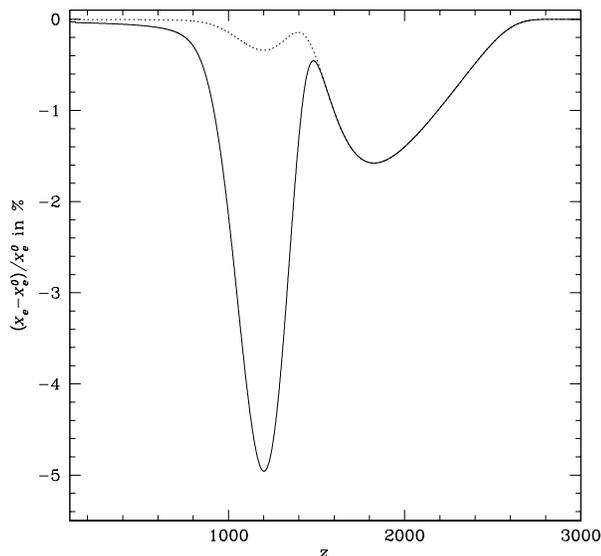}
\caption{The fractional difference (`new' minus `old') in $x_{\rm e}$ 
between the two models plotted in Fig.~\ref{graphxe}
as a function of redshift $z$.  The solid and dotted lines are the models
with the two-photon rates for H given by~\citet{dub05} and the scaled one given
by equation~(6), respectively.  Both curves are calculated using all the 
He\,{\small I} forbidden transitions as discussed in the text.}
\label{gdxe}
\end{figure}
The recombination histories calculated using the previous multi-level
code~\citep{seager00} and the code in this paper are shown
in Fig.~\ref{graphxe}, where $x_{\rm e} \equiv n_{\rm e}/ n_{\rm H}$
is the ionization fraction relative to hydrogen.  As we have included 
more transitions in our model, and these give electrons more 
channels to cascade down to the ground state, we expect the 
overall recombination rate to speed up, and that this will be noticeable 
if the rates of the extra frobidden transitions are significant.  
From Fig.~\ref{graphxe}, we can see that the recombination to 
He\,{\small I} is discernibly faster in the new calculation.  Fig.~\ref{gdxe}
shows the difference in $x_{\rm e}$ with and without the extra 
forbidden transitions.
The dip at around $z=1800$ corresponds to the recombination
of He\,{\small I} and the one around $z=1200$ is for H.  
Overall, the addition of the forbidden transitions claimed by~\citet{dub05}
 leads to greater than 1 per cent change in $x_{\rm e}$ over the redshift range where
the CMB photons are last scattering. 

\begin{figure}
\centering
\vspace*{7cm}
\leavevmode
\includegraphics{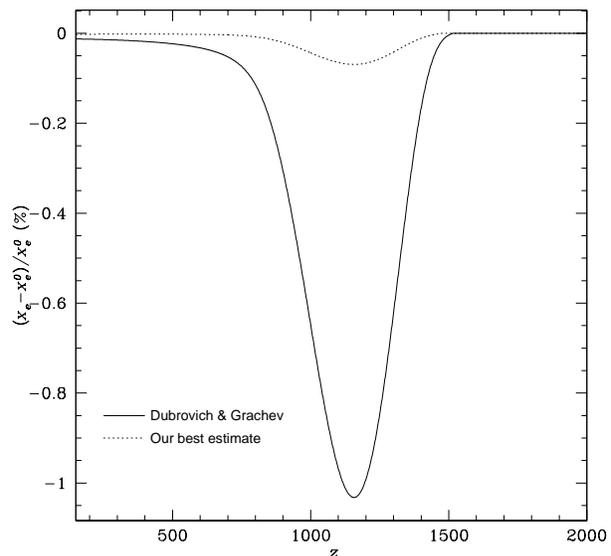}
\caption{Fractional change in $x_{\rm e}$ with the addition of the two-photon
transition from 3S and 3D to 1S for H.  The solid line is calculated with
the approximate rate given by~\citet{dub05} while the dashed line is 
calculated with the numerical rates given by~\citet{ctsc86}.} 
\label{Hn3}
\end{figure}

\begin{figure}
\centering
\vspace*{7cm}
\leavevmode
\includegraphics{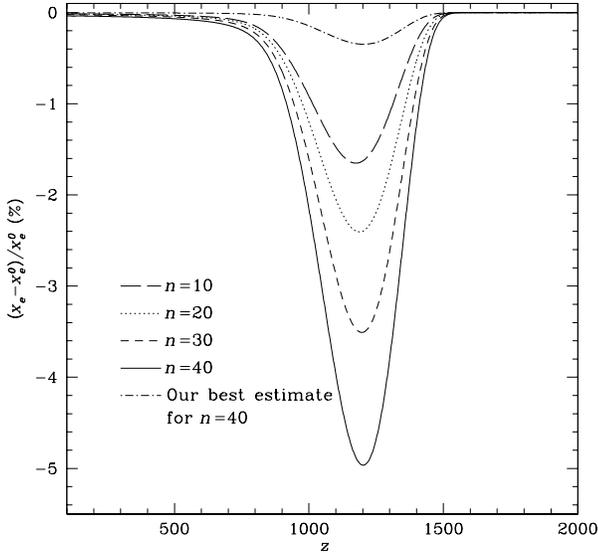}
\caption{Fractional change in $x_{\rm e}$ with the 
addition of different forbidden transitions for H.  The long-dashed, dotted,
dashed and solid lines include the two-photon transitions up to $n=10$, 20,
30 and 40, respectively, using the approximation for the rates given by equation~(5).
The dot-dashed line is calculated with the scaled rate from equation~(6).}
\label{gdxeH}
\end{figure}

In the last Section, we found that the approximated two-photon rate 
given by~\citet{dub05} for H with $n$\,$=$\,3 was overestimated by more 
than a factor of 10.  By considering only this extra two-photon transition,
 the approximate rate gives more than a per cent difference in $x_{\rm e}$, 
while with the more accurate numerical rates, the
change in $x_{\rm e}$ is less than 0.1 per cent~(as shown in Fig.~\ref{Hn3}).  Based on
this result, we do not need to include this two-photon transition, as it brings
much less than a per cent effect on $x_{\rm e}$.  For estimating the effect of
the extra two-photon transitions for higher $n$, we use the scaled
two-photon rate given by equation~(6). The result is plotted
in Fig.~\ref{gdxeH}.  The change in $x_{\rm e}$ with
the scaled two-photon rates is no more than 0.4 per cent, while the one
with the \citet{dub05} approximated rates brings about a 5 per cent change.

For He\,{\small I}, \citet{dub05} included the two-photon transitions 
from $n$\,$=$\,6 to 40, since they claimed that the approximate 
formula (equation~7) is good for $n$\,$>$\,6.  In our calculation, we use  
$\Lambda^{\rm HeI}_{n {\rm S}/n {\rm D}}$ from the approximate formula
 for the two-photon transitions of $n$\,$=$\,3 to 10, since this is the best one 
can do for now (and the formula at least gives the right order of magnitude).
The addition of the singlet-triplet $2^3{\rm P}_1$--$1^1{\rm S}_0$
transition and the $n^1{\rm S}_0$--$1^1{\rm S}_0$ and $n^1{\rm D}_2$--$1^1{\rm S}_0$ 
two-photon transitions with \mbox{$n$\,$=$\,3$-$10} cause more than
1 per cent changes in $x_{\rm e}$ (as shown in Fig.~2).  The 
$2^3{\rm P}_1$--$1^1{\rm S}_0$ transition has the biggest 
effect on $x_{\rm e}$.  

\begin{figure}
\centering
\vspace*{7cm}
\leavevmode
\includegraphics{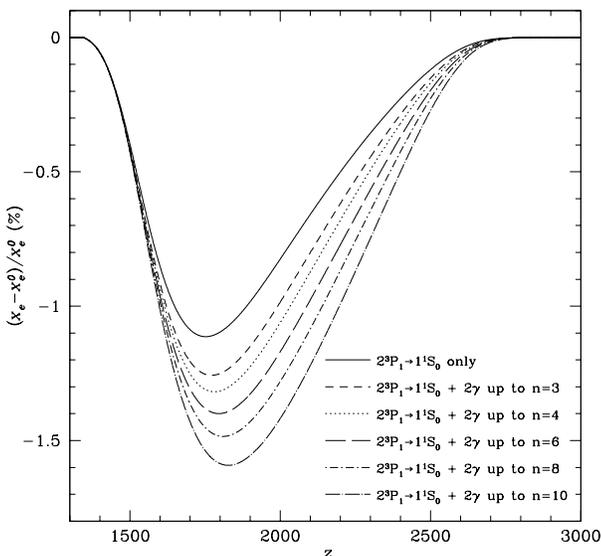}
\caption{Fractional change in $x_{\rm e}$ with the 
addition of different forbidden transitions for He\,{\small I} as a function 
of redshift.  The solid line corresponds to the calculation with only the
$2^3$P$_1$--$1^1$S$_0$ spin-forbidden transition.
The short-dashed, dotted, long-dashed, dot-dashed and long dot-dashed lines 
include both the spin-forbidden transition and the two-photon ($2 \gamma$) 
transition(s) up to $n=3, 4$, 6, 8 and 10, respectively.}
\label{gdxeHeI}
\end{figure}

\begin{figure}
\centering
\vspace*{7cm}
\leavevmode
\includegraphics{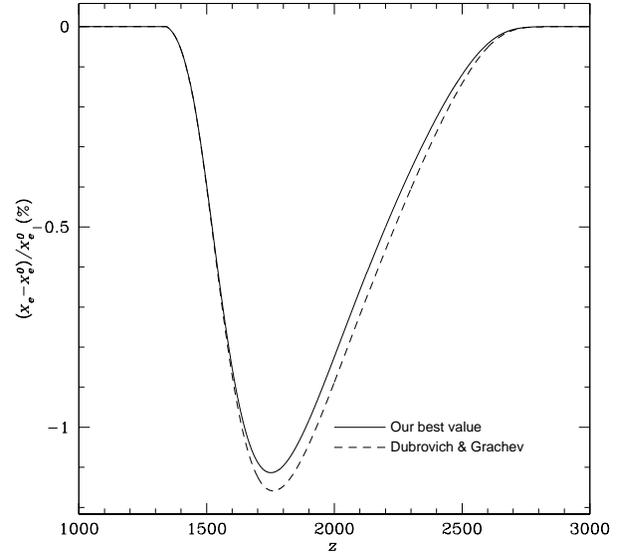}
\caption{Fractional change in $x_{\rm e}$ with only the 
$2^3$P$_1$--$1^1$S$_0$ forbidden transition.  The solid 
line is calculated with the rate $A=233$s$^{-1}$ from~\citet{dub05} and the 
dashed line is computed with our best value $A=177.58$s$^{-1}$
 from~\citet{lach01} .}
\label{gdxe23P}
\end{figure}

Fig.~\ref{gdxeHeI}  shows the fractional difference
in $x_{\rm e}$ using different combinations of additional forbidden transitions.
We can see that the $2^3{\rm P}_1$--$1^1{\rm S}_0$ transition
alone causes more than a 1 per cent change in $x_{\rm e}$, and the addition 
of each two-photon transition only gives about another 0.1 per cent change.  The extra
two-photon transitions from higher excited states (larger $n$) have a lower
effect on $x_{\rm e}$ compared with that from small $n$, and we 
checked that this trend continues to higher $n$.  However, 
the convergence is slow with increasing $n$.  Therefore,
one should also consider these two-photon transitions with 
\mbox{$n$\,$>$\,10} for He\,{\small I}, and the precise result will 
require the use of accurate rates, rather than an approximate 
formula such as equation~(7).  For the $2^3{\rm P}_1$--$1^1{\rm S}_0$
transition, \citet{dub05} adopted an older and slightly larger rate, 
and this causes a larger change of the ionization fraction 
(about 0.5 per cent more compared with that calculated with our 
best rate), as shown in Fig.~\ref{gdxe23P}.
%
%
%
\subsection{The importance of the forbidden transitions}
\begin{table*}
\centering
\begin{minipage}{120mm}
\caption{The percentage of electrons cascading down in each channel from 
$n=2$ states to the $1^1$S$_0$ ground state for He\,{\small I}.}
\begin{tabular}{@{}cccc@{}}
\hline
& $2^1$S$_0 \rightarrow 1^1$S$_0 $ & $2^1$P$_1 \rightarrow 1^1$S$_0 $ & 
$2^3$P$_1 \rightarrow1^1$S$_0$ \\
& (two-photon) & (resonant) & (spin-forbidden) \\
code from \citet{seager00} & 30.9\% & 69.1\% & -- \\
this work & 17.3\% & 39.9\% & 42.8\% \\ 
\hline
\end{tabular}
\end{minipage}
\end{table*}
\begin{figure}
\centering
\vspace*{7cm}
\leavevmode
\includegraphics{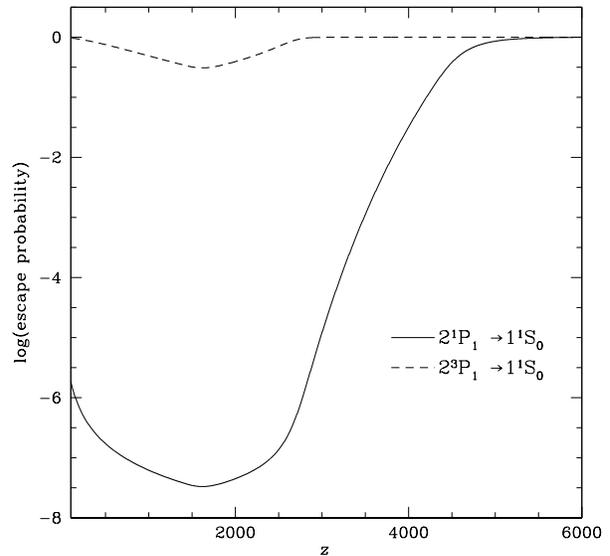}
\caption{Escape probability $p_{ij}$ as a function of redshift.
The solid line corresponds to the resonant transition between 
$2^1$P$_1$ and $1^1$S$_0$,
while the dashed line refers to the spin-forbidden transition between
$2^3$P$_1$ and $1^1$S$_0$.}
\label{gescp}
\end{figure}
%
\begin{figure}
\centering
\vspace*{7cm}
\leavevmode
\includegraphics{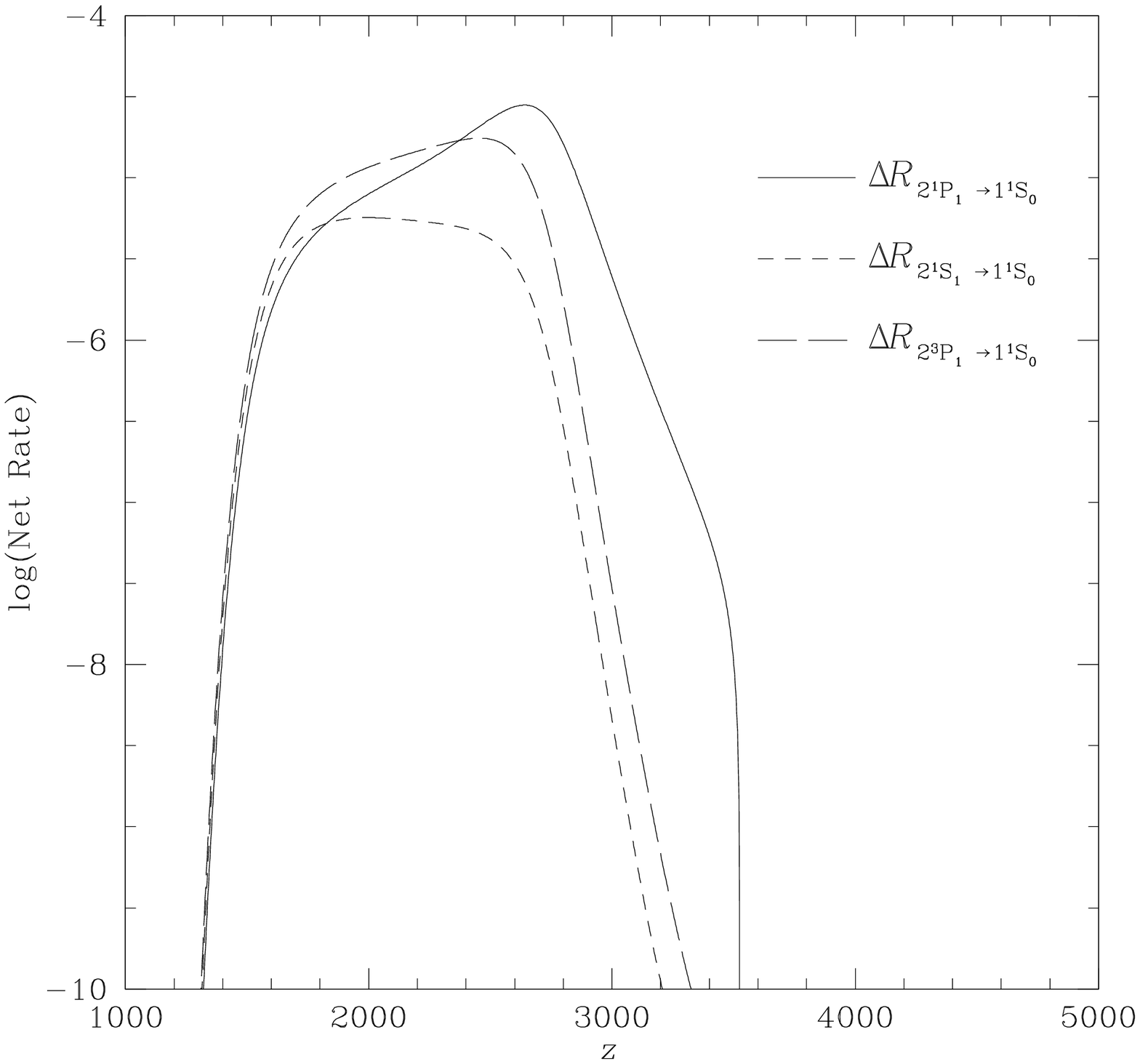}
\caption{Net bound-bound rates for He\,{\small I} as a function of redshift.
The solid line is the resonant transition between $2^1$P$_1$ and $1^1$S$_0$,
the short-dashed line is the two-photon transition between $2^1$S$_1$ and $1^1$S$_0$.
And the long-dashed line is the spin-forbidden transition between
$2^3$P$_1$ and $1^1$S$_0$.}
\label{grateHeI}
\end{figure}
%
One might wonder why the semi-forbidden transitions are significant
in recombination {\it at all}, since the spontaneous rate (or the Einstein $A$
coefficient) of the semi-forbidden transitions are about 6 orders of 
magnitude (a factor of $\alpha^2$, where $\alpha$ is the fine-sturcture 
constant) smaller than those of the 
resonant transitions. Let us take He\,{\small I} as an example for explaining 
the importance of the spin-forbidden $2^3{\rm P}_1$--$1^1{\rm S}_0$
transition in recombination.  The spontaneous rate 
is equal to 177.58\,s$^{-1}$ for this semi-forbidden transition, which is much
smaller than $1.7989 \times 10^9$\,s$^{-1}$ for the 
$2^1{\rm P}_1$--$1^1{\rm S}_0$ resonant transition.  But when we 
calculate the net rate [see equation~(8)], we also need to include 
the effect of absorption of the emitted photons by the surrounding 
neutral atoms, and we take this into account by multiplying the net 
bound-bound rate by the Sobolev escape probability $p_{ij}$~\citep{seager00}.  
If $p_{ij}$\,=\,1, the emitted line photons can 
escape to infinity, while if $p_{ij}$\,=\,0 the photons
will all be reabsorbed and the line is optically thick.  
Fig.~\ref{gescp} shows that the escape probability of the 
$2^1{\rm P}_1$--$1^1{\rm S}_0$ resonant transition is about 7 orders
of magnitude smaller than the spin-forbidden transition.  This makes the 
two net rates roughly comparable, as shown in Fig.~\ref{grateHeI}. 
From equation~(11), we can see that the easier it is to emit a photon, the 
easier that photon can be re-absorbed, because the optical depth $\tau_{\rm s}$ 
is directly proportional to the Einstein $A$ coefficient.  So when radiative 
effects dominate, it is actually natural to expect that some forbidden 
transitions might be important (although this is not true in a regime where 
collisonal rates dominate which is often the case in astrophysics).  
In fact for today's standard cosmological
model, slightly more than half of all the hydrogen atoms in the Universe
recombined via a forbidden transition~\citep{wong06}.  Table~1 shows 
that this is also true for helium.

In the previous multi-level calculation~\citep{seager00}, there was no 
direct transition between the singlet and triplet states.  The only 
communication between them was via the continuum, through the 
photo-ionization and photo-recombination transitions.  
Table~1 shows how many electrons cascade down through each channel from $n=2$ 
states to the ground state.  In the previous calculation, about 70\% of the 
electrons went down through the $2^1{\rm P}_1$--$1^1{\rm S}_0$ resonant transition.
In the new calculation, including the spin-forbidden transition between the 
triplets and singlets, there are approximately the same fraction 
of electrons going from the $2^1{\rm P}_1$ and $2^3{\rm P}_1$ states
to the ground state~(actually slightly more going from $2^3$P$_1$ 
in the current cosmological model).  This shows that we should certainly 
include this forbidden transition in future calculations. Our estimate
is that only about 40\% of helium atoms reach the ground state without
going through a forbidden transition. 

How about the effect of other forbidden transitions in He\,{\small I}
recombination?  We have included all the semi-forbidden electric-dipole
transitions with $n$\,$\leq$\,$10$ and $\ell$\,$\leq$\,$7$, and with
 oscillator strengths larger than $10^{-6}$ given by 
\citet{drake06}~(private communication).  
There is no significant change found in the ionization fraction. Besides 
the $2^3$P$_1$--$1^1{\rm S}_0$ transition, all the other extra 
semi-forbidden transitions are among the higher excited 
states where the resonant transitions dominate.  This is because these 
transition lines are optically thin and the escape probabilities are close 
to 1. 
\subsection{Effects on the anisotropy power spectrum}
\begin{figure}
\centering
\vspace*{7cm}
\leavevmode
\includegraphics{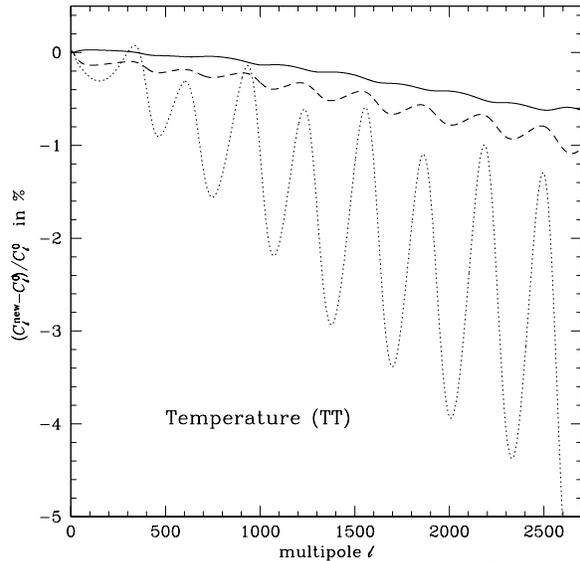}
\caption{Relative change in the temperature ({\sl TT}\/) angular power spectrum
due to the addition of the forbidden transitions.  The solid line 
includes the spin-forbidden transition and also the two-photons transitions
up to $n=10$ for He\,{\small I}, the dotted line includes all the 
above transitions and also the two-photon transitions up to $n=40$ for H
calculated with the approximate rates given by~\citet{dub05}. 
The dashed line is computed with the same forbidden transitions as the dotted 
line, but with our scaled rates~(and represents our best current estimate).}
\label{gclTT}
\end{figure}
\begin{figure}
\centering
\vspace*{7cm}
\leavevmode
\includegraphics{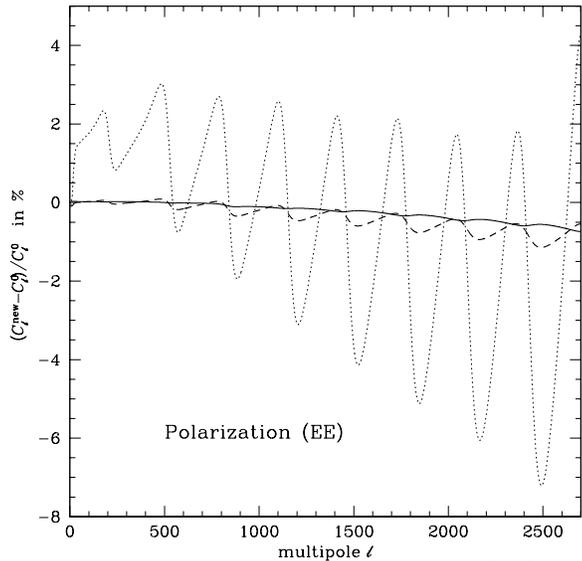}
\caption{Relative change in the polarization ({\sl EE}\/) angular power spectrum
due to the addition of the forbidden transitions, with the curves
the same as in Fig.~\ref{gclTT}.}
\label{gclEE}
\end{figure}

The CMB anisotropy power spectrum $C_{\ell}$ depends on the 
detailed profile of the evolution of the ionization fraction 
$x_{\rm e}$. This determines the thickness of the photon
last scattering surface, through the visibility function 
\mbox{$g(z) \equiv e^{- \tau} d \tau/dz$}, where $\tau$ is the 
Thomson scattering optical depth 
\mbox{($\tau = c\, \sigma_{\rm T} \int n_{\rm e} (dt/dz)\,dz$)}. 
The function $x_{\rm e}(z)$ sets the epoch when the tight coupling 
between baryons and photons breaks down, i.e. when the photon 
diffusion length becomes long, and the visibility 
function fixes the time when the fluctuations are effectively frozen 
in~\citep[see][and references therein]{hu95, seager00}.  
The addition of the extra forbidden transitions
speeds up both the recombination of H and He~{\small I}, and hence
we expect that there will be changes in $C_\ell$.

In order to perform the required calculation,
we have used the code {\sc cmbfast}~\citep{cmbfast} and modified it  
to allow the input of an arbitrary recombination history. 
Figs.~\ref{gclTT} and \ref{gclEE} show the relative changes in 
the CMB temperature ({\sl TT}\/) and polarization ({\sl EE}\/) anisotropy 
spectra, respectively, with different combinations of extra 
forbidden transitions.  The overall decrease of free electrons 
brings a suppression of $C_\ell$ over a wide range of $\ell$. 

For He\,{\small I}, there is 
less $x_{\rm e}$ at $z$\,$\simeq$\,$1400-2500$, which leads to an 
ealier relaxation of tight coupling.  Therefore, both the
photon mean free path and the diffusion length are longer.  Moreover,
the decrease of $x_{\rm e}$ in the high-$z$ tail results in increased
damping, since the effective damping scale is an average over the 
visibility function.  This larger damping scale leads to  
suppression of the high-$\ell$ part of the power spectrum.  From 
Figs.~\ref{gclTT} and \ref{gclEE}, we can see a decrease of 
$C_\ell$ (for both {\sl TT} and {\sl EE}) toward high $\ell$ 
for He\,{\small I}, with the maximum change being about 0.6 percent.

For H, the change of $C_\ell$ is due to the decrease in $x_{\rm e}$
at $z$\,$\simeq$\,$600-1400$ (see Fig.~\ref{gdxe}).  
There are two basic features in the curve of change in $C_{\ell}$ 
(the dotted and dashed lines in Fig.~\ref{gclTT}). Firstly, the power
spectrum is suppressed with increasing $\ell$, due to the lower 
$x_{\rm e}$ in the high-$z$ tail ($z$\,$>$\,1000). 
Secondly, there are a series of wiggles, showing that the 
locations of the acoustic peaks are slightly shifted.  This is due to 
the change in the time of generation of the $C_\ell$s in 
the low-$z$ tail.  $C^{\sl EE}_\ell$ actually shows an increase 
for $\ell \leq 1000$ (see Fig.~\ref{gclEE});
this is caused by the shift of the center of the visibility 
function to higher $z$, leading to a longer diffusion length.
Polarization occurs when the anisotropic hot and cold photons are 
scattered by the electrons.  The hot and cold photons can 
interact with each other within the diffusion length, and therefore,
a longer diffusion length allows more scatterings and leads to a 
higher intensity of polarization at large scales.  

With the approximate rates used by~\citet{dub05}, the maximum relative 
change of $C^{\sl TT}_\ell$ is about 4 percent and for 
$C^{\sl EE}_\ell$ it is about 6 percent.  
The overall change is thus more than 1 percent over a wide 
range of $\ell$.  However, if we adopt the scaled two-photon rate given
by equation~(6), the relative changes of $C^{\sl TT}_\ell$ and 
$C^{\sl EE}_\ell$ are no more than 1 per cent.  Note that
we do not plot the temperature-polarization 
correlation power spectrum here, since there is no dramatically
different change found (and relative differences are less meaningful 
since $C^{\sl TE}_\ell$ oscillates around zero).
%
\section{Discussions}
In our model we only consider the semi-forbidden transitions with 
$n$\,$\leq$\,10 and $\ell$\,$\leq$\,7 for He\,{\small I} and 
the two-photon transitions from the higher S and D states to the ground
state for H and He\,{\small I}.  It would be desirable to perform 
a more detailed investigation of all the other forbidden transitions, which may 
provide more paths for the electrons to cascade down to the ground state
and speed up the recombination process.  
In this paper we have tried to focus on the forbidden transitions 
which are likely to be the most significant.  However we caution that, if the 
approximations used are inadequate, or other transitions prove to be 
important, then our results will not be accurate.

There are several other approximations that we have adopted in order
to perform our calculations.  For example,
we consider the {\it non-resonant} two-photon rates for
higher excited rates.  The two-photon transitions from higher excited 
states ($n$\,$\ge$\,3) to the ground state are more complicated than 
the 2S--1S transition, because of the resonant intermediate states.
For example, for the 3S--1S two-photon transition, the spectral 
distribution of the emitted photons shows infinities (resonance
peaks) at the frequencies corresponding to the 3S--2P and 2P--1S
transitions~\citep{tung84}.  Here, we use only the non-resonant
rates, by assuming a smooth spectral distribution of the 
emitted photons; this probably gives a lower limit on the change of
$x_{\rm e}$ and $C_\ell$ coming from these extra forbidden transitions.
The correct way to treat this would be to consider the rates and 
feedback from medium using the full spectral distribution of the 
photons and radiative transfer;  this will have to wait for a
future study.

Besides the consideration of more forbidden transitions, there are 
many other improvements that could be made to the recombination 
calculation.  In particular,
\citet{rub06} showed that a multi-level calculation of
the recombination of H with the inclusion of seperate $\ell$-states
can give more than 20 per cent difference in the population 
of some levels compared with the thermal equilibrium assumption
for each $n$-shell.  The latest calculation, considering up to 100 
shells, is presented by~\citet{CRMS06}, but does not include
all the forbidden transitions studied here. A more complete calculation
 should be done by combining the forbidden transitions in a code with full
angular momentum states, and we leave this to a future study.  There
are also other elaborations which could be included in future
calculations, which we now describe briefly.

The rate equation we use for all the two-photon transitions only
includes the spontaneous term, assuming there is no interaction
with the radiation background (see equation~(3)).  \citet{chluba05}
suggested that one should also consider the stimulated effect
of the 2S--1S two-photon transition for H, due to photons 
in the low frequency tail of the CMB blackbody spectrum.  
\citet{leung04} additionally argued that the change of the adiabatic 
index of the matter should also be included, arising due to the neutralization
of the ionized gas.  These two modifications have been studied only 
in an effective three-level atom model, and more than a percent change
in $x_{\rm e}$ was claimed in each case~(but see~\citet{wong06b} 
for arguments against the effect claimed by Leung et al.~2004).

For the background radiation field $\bar{J}$, we approximated it
with a perfect blackbody Planck spectrum.  This approximation
is not completely correct for the recombination of H, since the He line 
distortion photons redshift into a frequency range that can in principle
photo-ionize the neutral H~\citep{dell93,seager00, wong06}.  
Althought we expect this secondary distortion effect to bring the 
smallest change on $x_{\rm e}$ among all the modifications suggested here, 
it is nevertheless important to carry out the calculation 
self-consistently, particularly for the spectral line distortions.
In order to obtain an accurate recombination history, 
we therefore need to perform a full multi-level calculation with 
seperate $\ell$-states and all the improvements suggested above,
which we plan to do in a later paper.

For completeness we also point out that the accuracy of the physical 
constants is important for recombination
as well.  The most uncertain physical quantity in the recombination
calculation is the gravitational constant $G$.  The value of $G$ used
previously in the
{\sc recfast} code is $6.67259$\,$\times$\,$10^{-11}$m$^{3}$kg$^{-1}$s$^{-2}$ and 
the latest value \citep[e.g. from the Particle Data Group,][]{pdg06} is 
$6.6742$\,$\times$\,$10^{-11}$m$^{3}$kg$^{-1}$s$^{-2}$.  Another quantity 
we need to modify is the atomic mass ratio of $^4$He and $^1$H, 
$m_{\rm ^4He}/m_{\rm ^1H}$, which was previously taken to be 
equal to 4\,\citep[as pointed out by][]{steigman06}.  
By using the atomic masses given by \citet{pdg06}, 
the mass ratio is equal to 3.9715.  The overall change in $x_{\rm e}$
is no more than 0.1 per cent after updating these two constants in both
{\sc recfast} and multi-level code.  

\section{Conclusions}
In this paper, we have computed the cosmological recombination 
history by using a multi-level code with the addition of the 
$2^3$P$_1$ to $1^1$S$_0$ spin-forbidden
transition for He\,{\small I} and the two-photon transitions 
from $n$S and $n$D states to the ground state for both
H and He\,{\small I}.  With the approximate rates from~\citet{dub05},
we find that there is more than a per cent decrease in the ionization 
fraction, which agrees broadly with the result they claimed.  
However, the only available accurate numerical value of two-photon rate 
with $n\geq 3$ is for the 3S to 1S and 3D to 1S transitons for H.  
We found that the approximate rates from~\citet{dub05} 
were overestimated, and instead we considered a scaled rate in order 
to agree with the numerical $n=3$ two-photon rate.  With this scaled 
rate, the change in $x_{\rm e}$ is no more than 0.5 per cent.

Including these extra forbidden transitions,
the change in the CMB anisotropy power spectrum is more 
than 1 per cent, which will potentially 
affect the determination of cosmological parameters in future
CMB experiments.  Since one would like the level of theoretical
uncertainty to be negligible, it is essential to include these
forbidden transitions in the recombination calculation.  In addition,
we still require accurate spontaneous rates to be calculated
for the two-photon transitions and also a code which includes
at least all the modifications suggested in Section~4, 
in order to obtain the $C_\ell$s down to the 1 per cent level.  
Achieving sub-percent accuracy in the calculations is challenging!

However, the stakes are high -- the determination of the parameters
which describe the entire Universe -- and so further work will be 
necessary.  Systematic deviations of the sort we have shown would 
potentially lead to incorrect values for the spectral tilt derived
from Planck and even more ambitions future CMB experiments, and 
hence incorrect inferences about the physics which produced the 
density perturbations in the very early Universe.  It is 
amusing that in order to understand physics at the $10^{15}$\,GeV 
energy scale we need to understand eV scale physics in exquisite 
detail!
\section{Acknowledgements}
This work would have been impossible without the earlier discussions 
with Sara Seager.
We thank Donald C. Morton for useful conversations and providing 
us with updated forbidden transition rates.  We are grateful to
Peter J. Storey for providing the data for the photo-ionization cross-section
of neutral helium.  We would also like to thank Alexander Dalgarno, 
Victor K. Dubrovich, Christopher Hirata and Eric Switzer for useful discussions. 
This work made use of the code {\sc cmbfast}, written by
Uros Seljak and Matias Zaldarriaga. 
This research was supported by the Natural Sciences and Engineering Research
Council of Canada and a University of British Columbia Graduate Fellowship.
%

\bsp

\label{lastpage}

\end{document}